\documentclass[twocolumn,secnumarabic,amssymb, floatfix, nobibnotes, aps, prl, superscriptaddress]{revtex4-1}
\pdfoutput=1
\usepackage{dcolumn} 
\usepackage{natbib}
\usepackage{bm} 
\usepackage{float} 
\usepackage{bbm} 
\usepackage{color}
\usepackage{amsfonts}
\usepackage{lmodern} 
\usepackage{amsmath,amssymb}
\usepackage[normalem]{ulem}
\usepackage{svg}
\usepackage{placeins}

\renewcommand{\vec}[1]{\boldsymbol{#1}}
\newcommand{\tens}[1]{\boldsymbol{#1}}
\newcommand{\bnabla}{\vec{\nabla}}

\begin{document}
\title{Cellular dynamics emerging from turbulent flows steered by active filaments}
\author{Mehrana R. Nejad}
\affiliation{Department of Physics, Harvard University, Cambridge, MA 02138}
\affiliation{School of Engineering and Applied Sciences, Harvard University, Cambridge, MA 02138, USA}
\email{mehrana@g.harvard.edu}
\author{Julia M. Yeomans}
\affiliation{The Rudolf Peierls Centre for Theoretical Physics, Department of Physics, University of Oxford, Parks Road, Oxford OX1 3PU, United Kingdom}
\author{Sumesh P Thampi}
\affiliation{The Rudolf Peierls Centre for Theoretical Physics, Department of Physics, University of Oxford, Parks Road, Oxford OX1 3PU, United Kingdom}
\affiliation{Department of Chemical Engineering, Indian Institute of Technology, Madras, Chennai, India 600036}


\begin{abstract}
We develop a continuum theory to describe the collective dynamics of deformable epithelial cells, using two tensor order parameters to distinguish the force-generating active filaments in the cells  from their shape. The theory demonstrates how active flows create nematic domains of anisotropic cells, drive active turbulence, and create topological defects. We show that the filament flow-aligning parameter, $\lambda_{\tens{Q}}$, a rheological quantity that determines the response of the filaments to velocity gradients in the active flows, plays a significant, to date unappreciated, role in determining the pattern of extensional and compressional active flows. In a contractile cell layer, local flows are expected to align elongated cells perpendicular to the active filaments. However, with increasing $\lambda_{\tens{Q}}$, long-range correlations in the active turbulent flow field lead to extended regions where this alignment is parallel, as recently observed in experiments on confluent MDCK cell layers. Further, the two order parameter formalism allows us to distinguish defects in the filament director field, which contribute to the active driving, and those in the shape director field, measured in experiments, which are advected by the active flows.  By considering the relative orientations of shape and filaments we are able to explain the surprising observation of defects moving towards their heads in contractile cell layers.
\end{abstract}

\maketitle
Active matter describes materials that operate out of thermodynamic equilibrium, harnessing energy from their surroundings to do work. A canonical example is living matter which transforms chemical energy to growth and movement, the hallmarks of life. Therefore there is considerable interest in the biophysics community in applying active matter physics to investigate mechanical and developmental processes in biology \cite{gompper20202020,hallatschek2023proliferating,bowick2022symmetry,yeomans2023mechanobiology,joanny2022tissues}. Examples include correlated tissue flows in embryogenesis \cite{gehrels2023curvature,herrera2023tissue}, wound healing \cite{tetley2019tissue, alert2020physical}, and cancer development and metastasis \cite{laang2024topology, grosser2021cell}. 

Epithelia are an important class of tissue, that include the skin and the lining of hollow organs such as the lungs. Epithelial cells often form tightly connected, two-dimensional layers, and well-studied in vitro model systems consist of confluent layers of epithelial cells cultured on a substrate \cite{lemke2021dynamic,villeneuve2024mechanical,ascione2022collective,saadaoui2020tensile,tetley2019tissue,duclos2018spontaneous}. The activity of the cells can result in complex, dynamical behaviour, including flocking \cite{saraswathibhatla2021coordinated, bowick2022symmetry} and active turbulence \cite{saw2017topological, alert2020universal}.  Recent studies have increasingly highlighted the importance of epithelial dynamics in organoids and in vivo \cite{lee2022epithelial,devany2021cell,cavanaugh2020rhoa}.
 
 The biological processes involved in collective cell motion are complex and not well understood, and the available experimental data is diverse. A  useful approach to address these challenges is to develop coarse-grained, continuum theories without cellular details, but based on conservation laws and symmetries. These theories provide a foundation for describing the mechanics of cell collectives and tissues.
A successful example is continuum theory of active nematics which has offered a framework to explain several properties of cell layers. These include the dynamical state of active turbulence, characterised by jets and vortices of the velocity field, and driven by 
non-equilibrium forces arising from stress generating active filaments \cite{prost2015active, saw2018biological, doostmohammadi2018active, Balasubramaniam2021}, cellular division \cite{joanny2023statistical, PhysRevLett.117.048102} or topological rearrangements \cite{shankar2022topological,xu2023geometrical}, the presence of topological defects that trigger cell extrusion \cite{saw2017topological, vafa2022active}, and motility and spontaneous oscillations that arise from self-generated flows \cite{duclos2018spontaneous,armengol2023epithelia, blanch2018turbulent,PRXLife.1.023008,ascione2022collective}. 

Almost all these studies assume that the microstructure of cell collectives can be described by a single tensor order parameter, the $\tens{Q}$ tensor, which represents the nematic order in the system. The literature variously interprets the $\tens{Q}$ tensor as referring to force generating active filaments or cellular shape. In this paper we show that it is imperative to distinguish the shape of the cells from the internal stress generating active filaments, and that two tensor order parameters are required to capture many features of the collective dynamical behaviour of cells.

In the next section  we describe the model. An active nematic field, representing active filaments, produces flows. A second tensor field describes cell shape which can be altered by the active flows. We show how the coupled equations of motion predict the possibility of locally extensional flows in a contractile active material, in agreement with recent experiments \cite{nejad2024stress}. By distinguishing topological defects in the filament and shape fields we explain why shape defects can move towards their heads or tails in a contractile system, thus resolving a long-standing puzzle in the literature \cite{balasubramaniam2021investigating,duclos2018spontaneous,vafa2021,killeen2022}.

\subsection*{Theory} We use a continuum approach and describe the dynamics of a two-dimensional layer of confluent cells using a vector velocity field $\vec{u}$ and two tensor order parameter fields for the microstructure $\tens{Q}$ and $\tens{W}$. The rank two tensor $\tens{Q}$, referred to as the filament order parameter, represents the nematic orientational order of the filaments that form the cytoskeletal network. As well studied in the active nematic literature, $\tens{Q}$ is a traceless, symmetric tensor; the eigenvalues and eigenvectors give the magnitude of the orientational order, $S$, and the director field respectively \cite{DeGennesBook, beris1994thermodynamics, Sriram2002}. Motor proteins bound to the cytoskelatal filaments generate internal stresses in the cellular layer using the energy derived from adenosine triphosphate (ATP) hydrolysis, as demonstrated both in vivo and in vitro experiments \cite{sanchez2012spontaneous, saw2017topological, duclos2018spontaneous}. We model these by including a term that generates active stresses and flows in the equations of motion \cite{colen2021machine,PhysRevLett.92.118101,Sriram2002}.

Active flows, cell-cell adhesion, cellular processes, and mechanical stress are known to affect the cell shape \cite{alert2020physical, luciano2022appreciating}, which may evolve independently from the dynamics of the cytoskeletal filamentous network \cite{nejad2024stress, schakenraad2020mechanical}.  To account for this,  we introduce a second structural variable  to describe cell shape, a quantity often characterized in experiments.  The shape order parameter $\tens{W}$ is a traceless, symmetric, rank two tensor that quantifies anisotropy in the cell shape. $\tens{W} = \tens{0}$ corresponds to isotropic cells and the eigenvalues and eigenvectors of $\tens{W}$ give the magnitude of the anisotropy of the cells and the direction of the corresponding deformation, respectively. 

The velocity field obeys the continuity and Navier-Stokes equations,
\begin{align}
\bnabla\cdot\vec{u} &=0, \label{eqn:comp}
\\
\rho \left(\partial_t + \vec{u}\cdot\bnabla\right) \vec{u} &= \bnabla\cdot\tens{\Pi}.\label{eqn:ns}
\end{align}
Here, $\rho$ is the local density and $\tens{\Pi}$ is a generalized stress tensor that contains contributions from active cellular processes and (passive) visco-elastic effects.

The two tensor order parameters are advected by the flow fields, and evolve in response to flow gradients and thermodynamic driving forces. Thus,
\begin{align}
\partial_t \tens{Q}=& - \vec{u}\cdot \boldsymbol{\nabla} \tens{Q} + \mathbf{S}_{\tens{Q}} + \Gamma_{\tens{Q}} \tens{H}_{\tens{Q}},\label{qdynamics}\\
\partial_t \tens{W}=& - \vec{u}\cdot \boldsymbol{\nabla} \tens{W} + \mathbf{S}_{\tens{W}} + \Gamma_{\tens{W}} \tens{H}_{\tens{W}},\label{ddynamics}
\end{align} 
where, for $\tens{R} \in \{\tens{Q}, \tens{W}\}$, $\partial_t {\tens{R}} + \vec{u}\cdot \boldsymbol{\nabla} \tens{R} - \mathbf{S}_{\tens{R}}$ is the convected derivative for the tensor order parameter field $\tens{R}$ and
\begin{align}
\mathbf{S}_{\tens{R}} &= (\lambda_{\tens{R}} \tens{E} + \boldsymbol{\Omega}) \cdot \left( \tens{R} + \frac{\tens{I}}{2} \right) + \left( \tens{R} + \frac{\tens{I}}{2} \right) \nonumber\\&\cdot(\lambda_{\tens{R}} \tens{E} - \boldsymbol{\Omega}) -\lambda_{\tens{R}} \left( \tens{R} + \frac{\tens{I}}{2} \right)(\tens{R} : \nabla \vec{u}).   
\end{align}
The strain rate tensor and the vorticity tensor are defined as $\tens{E}=(\bnabla\vec{u}^{T}+\bnabla\vec{u})/2$ and $\tens{\Omega}=(\bnabla\vec{u}^{T}-\bnabla\vec{u})/2$, respectively, representing the symmetric and antisymmetric parts of the velocity gradient tensor.
\begin{figure*}[ht] 
    \centering
    \includegraphics[width=\linewidth]{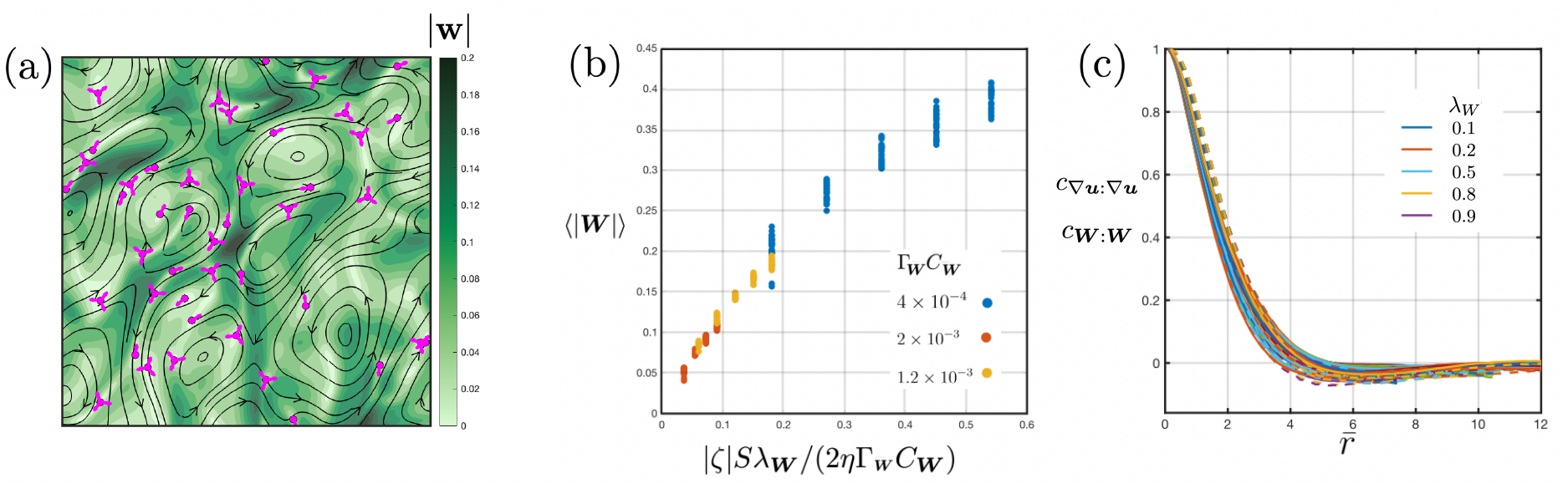}
    \caption{{\bf Active flows form elongated, nematically ordered cells.} (a) Active turbulence illustrated using the velocity streamlines. The domain is colored with cell elongation $|\tens{W}|$. Comet-shaped and trefoil-shaped symbols are $\pm\frac{1}{2}$ topological defects in the nematic order of the cell shape field. (b) $\langle|\tens{W}|\rangle$, spatio-temporally averaged magnitude of cell elongation, as a function of contractile activity $|\zeta|$ of the filaments. Here, $\lambda_{\tens{Q}}=0$, $\lambda_{\tens{W}}=1$. 
    (c) Spatial correlations in the shape tensor $c_{\tens{W}:\tens{W}} (r)$ (continuous lines) and the velocity gradient tensor $c_{\nabla\vec{u} :\nabla\vec{u}}$ (dashed lines) as a function of distance $\bar{r}= r/l_a$ which is normalised by the active length scale $l_a= \sqrt{K_{\tens{Q}}/|\zeta|}$, for various values of activity, $\zeta \in \{\pm 0.01, \pm 0.02, \pm 0.03\}$. The correlation functions are defined as $c_{\tens{X}:\tens{X}} (r)= (\langle \tens{X}(\textbf{r}_0) : \tens{X}(\textbf{r}_0+\textbf{r})\rangle)/(\langle \tens{X}(\textbf{r}_0) : \tens{X}(\textbf{r}_0)\rangle)$, where $\tens{X} \in \{\boldsymbol{\nabla} \boldsymbol{u} , \boldsymbol{W}\}$ and $\langle \rangle$ shows the average over space $r_0$ and over time. 
    }
    \label{fig1fi}
\end{figure*}

Different weights are given to the symmetric and antisymmetric parts of the velocity gradient tensor in coupling with the field $\tens{R}$ through the parameter $\lambda_{\tens{R}}$. The filament flow-aligning parameter $\lambda_{\tens{Q}}$, also known as flow-tumbling parameter in the liquid crystal literature, is a rheological quantity that is related to the Leslie coefficients of anisotropic viscosity \cite{beris1994thermodynamics}. For nematogens with fixed shape, this parameter is a function of aspect ratio and orientational order, and it determines the aligning/tumbling behavior of active filaments in a simple shear flow.  Analogously, we have $\lambda_{\tens{W}}$ as a shape flow-aligning parameter. For deformable particles, this coefficient is positive and leads to the elongation of the particles along the principal directions of the velocity gradient tensor. In the limit $\lambda_{\tens{W}} \to 0$ flow gradients merely rotate the tensor field $\tens{W}$, that is, cells do not undergo any deformation but they undergo rigid body rotation in accordance with the local vorticity tensor. We consider $\lambda_{\tens{W}} > 0$ in this study.

Further,  $\Gamma_{\tens{R}}$ is the orientational diffusivity and
\begin{align}
   \tens{H}_{\tens{R}} = -\frac{\delta \mathcal{F}_{\tens{R}}}{\delta \tens{R}} + \frac{\tens{I}}{2}\textnormal{Tr} \left(\frac{\delta \mathcal{F}_{\tens{R}}}{\delta \tens{R}} \right)
\end{align} 
is the molecular potential derived from the variational derivative of the free energy $\mathcal{F}_{\tens{R}} = \int f_{\tens{R}} d\mathcal{V}$, where $\tens{I}$ is the identity tensor and $Tr(\cdot)$ is the trace of the tensor argument. The free energy density $f_{\tens{R}}$ is 
\begin{align}\label{ela}
f_{\tens{R}} = f^{b}_{\tens{R}}
 + \frac{K_{\tens{R}}}{2} |\nabla \tens{R}|^2, 
\end{align}
where the bulk terms are chosen as $f^{b}_{\tens{Q}} = \frac{C_{\tens{Q}}}{4} (1- 3\tens{Q} : \tens{Q})^2 $ and $f^{b}_{\tens{W}} = \frac{C_{\tens{W}}}{4} (1 + \tens{W} : \tens{W})^2 $.
$f^{b}_{\tens{Q}}$ corresponds to the usual Landau-de Gennes free energy contribution for the $\tens{Q}$ tensor field \cite{DeGennesBook} while $f^{b}_{\tens{W}}$ ensures that the shape tensor corresponds to isotropic cells in the absence of activity.  The second term in Eq.~\ref{ela} 
determines the contribution arising from the gradients in the field variable (Frank elasticity) assuming a single elastic constant, $K_{\tens{R}}$.

While the spatio-temporal evolution of the two order parameters $\tens{Q}$ and $\tens{W}$ is slaved to the flow field $\vec{u}$, the two microstructural variables also dictate the dynamics of the cellular layer through the stress, $\tens{\Pi}$, in Eq.~\ref{eqn:ns}. The stress field includes both active and passive contributions. The active part, $ \tens{\Pi}^\textmd{a} = -\zeta \textbf{Q}$, arises from the internal stress generated by the action of molecular motors on the cytoskeletal filaments. The passive part includes the isotropic contribution $ \tens{\Pi}^\textmd{i} = -P\tens{I}$, where $P$ is the pressure, the viscous stress, $\tens{\Pi}^\textmd{v} = 2 \eta \tens{E}$, where $\eta$ is the medium viscosity, and 
elastic stresses arising from the two tensor order parameters, 
\begin{align}
\tens{\Pi^{\text{e}}}_{\tens{R}} &= 2 \lambda_{\tens{R}} \tens{{R}} (\tens{R}:\tens{H}_{\tens{R}}) -\lambda_{\tens{R}}( \tens{H}_{\tens{R}}\cdot\tens{{R}}  + \tens{{R}} \cdot \tens{H}_{\tens{R}}) \nonumber\\
& -\tens{\nabla}\tens{R} : \frac{\delta \mathcal{F}_{\tens{R}}}{\delta \tens{\nabla}\tens{R}} + \tens{R}\cdot\tens{H}_{\tens{R}} - \tens{H}_{\tens{R}}\cdot\tens{R}.
\label{eqn:elastic}
\end{align}

\begin{figure*}[ht] 
    \centering
    \includegraphics[width=\linewidth]{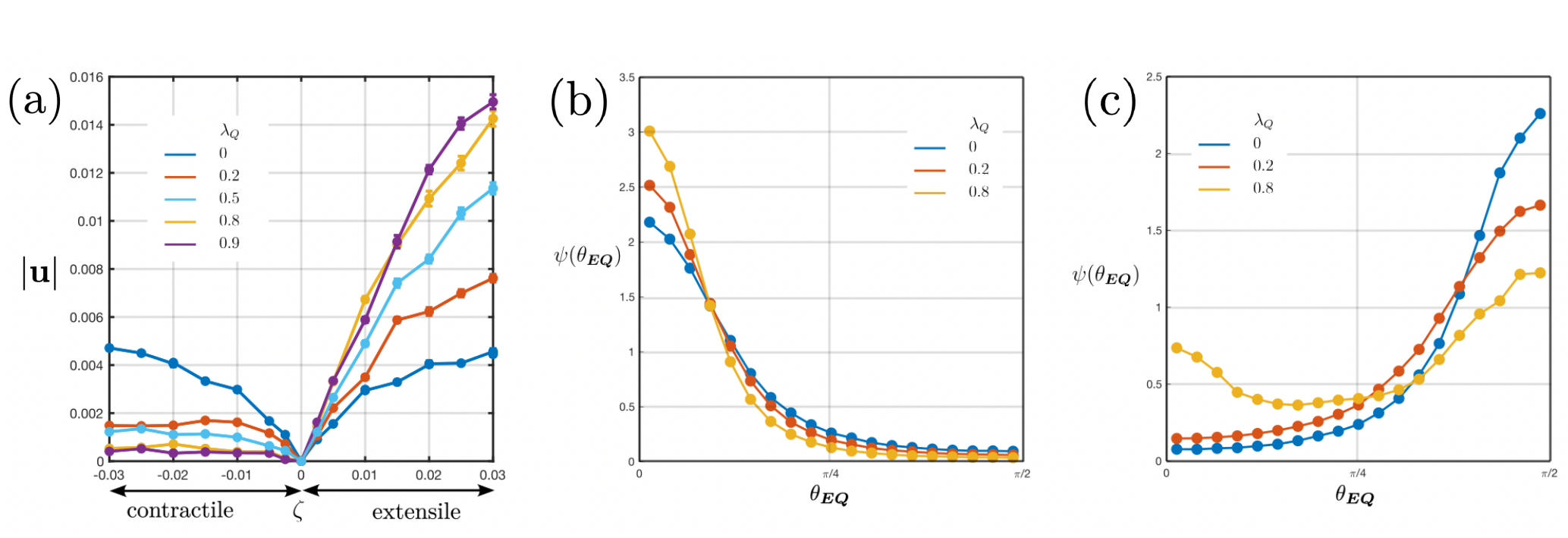}
   \caption{{\bf Dependence of the active turbulent flow field on the filament flow-aligning parameter $\lambda_{\tens{Q}}$.} (a) Average speed of the fluid as a function of filament activity. (b,c) Probability density function $\psi(\theta_{\tens{E} \tens{Q}})$ describing the distribution of $\theta_{\tens{E}\tens{Q}}$, the angle between the dominant eigenvectors of the rate of strain tensor and the filament order parameter in an active turbulent flow field in (b) extensile and (c) contractile systems.  $\psi(\theta_{\tens{E} \tens{Q}})$ is normalized so that the area under the graph is equal to 1.
}\label{fig2fi}
\end{figure*}
\subsection*{Simulations}
We use a hybrid lattice Boltzmann method \cite{PhysRevE.76.031921,thampi2014vorticity} to solve the set of governing equations \ref{eqn:comp}--\ref{ddynamics}. The simulations were initialized with a quiescent fluid ($\vec{u} = 0$), with randomly oriented active filaments ($\tens{Q}$) and shapes ($\tens{W}$) in a domain of $150 \times 150$ with periodic boundary conditions. Data was collected after the simulations reached a statistically steady state. Unless otherwise stated the parameters used in the simulations were: $C_{\tens{Q}} =C_{\tens{W}} = 10^{-3}$, $K_{\tens{Q}} = 10^{-2}$, $K_{\tens{W}} = 0$, $\Gamma_{\tens{Q}} = \Gamma_{\tens{W}} = 0.4$, $\rho=40$, $\eta = 1$ and $\zeta=\pm0.01 $. 

The active stress $\tens{\Pi}^{a}$ that arises from the activity of molecular motors on the filaments results in the well known hydrodynamic instabilities of active nematics. Consequently, the state of active turbulence prevails in the system. The turbulent state is sustained through the formation and destruction of domain walls and topological defects in the nematically ordered filaments \cite{head2024spontaneous, thampi2014instabilities, giomi2013defect}.

\subsection*{Cells are elongated by active flows}

The active turbulent flow field consists of fluid jets and vortices \cite{blanch2018turbulent, thampi2014vorticity, giomi2014defect} as shown in Fig.~\ref{fig1fi}(a). The straining flows of active turbulence in a confluent cell layer deform the otherwise isotropic cells. The shading in the figure shows the magnitude of the shape tensor field, $|\tens{W}|$, indicating the extent of this deformation. Further, Fig.~\ref{fig1fi}(b) plots the anisotropy of the cells, quantified as the average magnitude of the shape tensor, $\langle|\tens{W}|\rangle$, as a function of the contractile activity of the filaments, $|\zeta|$.

Examining Eq.~\ref{ddynamics} for a homogeneous steady-state solution for the shape tensor, for small deformations ($|\tens{W}|\ll1$) we obtain
\begin{align}
\tens{W} \sim \frac{\lambda_{\tens{W}}}{\Gamma_{\tens{W}} C_{\tens{W}} } \tens{E},
\label{eq:WErelation}
\end{align}
confirming that the straining flows of active turbulence indeed deform the cells with the direction of deformation  along the principal axis of the strain rate tensor, $\tens{E}$. Balancing the viscous and active stress in the momentum balance (Eq.~\ref{eqn:ns}) gives
\begin{align}
    \tens{E} \sim \frac{\zeta}{2 \eta} \tens{Q}.
    \label{eq:EQrelation}
\end{align}  
Eliminating $\tens{E}$ from Eq.~\ref{eq:WErelation} using Eq.~\ref{eq:EQrelation} we obtain a linear scaling for the deformation of the cells $ |\tens{W}| \sim |\zeta|  \lambda_{\tens{W}} S /(2\eta \Gamma_{\tens{W}} C_{\tens{W}})$ consistent with the data reported in Fig.~\ref{fig1fi}(b). At higher activities, the scaling breaks down since the elongation of cells is not small and the contribution of nonlinear terms in the free energy (Eq.~\ref{ela}) is not negligible.  

In order to probe whether the anisotropic cells develop spatial nematic order, even given our choice of elastic constant $K_{\tens{W}}=0$, we measure the spatial correlations of the shape tensor, $\langle \tens{W}(\mathbf{r}_0) : \tens{W}(\mathbf{r}_0+\mathbf{r})\rangle$, normalised by $\langle |\tens{W}(\mathbf{r}_0)|^2\rangle$ and plotted in Fig.~\ref{fig1fi}(c). The spatial correlations in the velocity gradient tensor, $\langle \nabla\vec{u}(\mathbf{r}_0):\nabla\vec{u}(\mathbf{r}_0+\mathbf{r})\rangle$, normalised by $\langle |\nabla\vec{u}(\mathbf{r}_0)|^2\rangle$ which decay over the active length scale $\sim\sqrt{K_{\tens{Q}}/|\zeta|}$, are also shown for comparison. The spatial nematic order of anisotropic cells persists on the active length scale, and behaves similarly to the spatial correlations in the velocity gradient field, as expected since the former is driven by the latter. Similarly, both shape deformations and flows persist over the active time scale, $\tau_a \sim \eta/\zeta$ \cite{velez2024probing} (see Fig.~\ref{fig:timecorr}).
Hence, activity alone can lead to both elongation and spatial nematic order of the elongated cells without the need for any thermodynamic driving forces.

While straining flows generated by activity elongate the deformable cells, the corresponding vortical field (antisymmetric part of the velocity gradient tensor)  simultaneously rotates them. Therefore, the filament flow-aligning parameter $\lambda_{\tens{Q}}$ and the shape flow-aligning parameter $\lambda_{\tens{W}}$ that control the response of the nematic order parameter of the filaments $\tens{Q}$ and that of the shape order parameter of the cells $\tens{W}$ to the strain rate and vorticity tensors play an important role in determining the deformation of the cells. Irrespective of the value of $\lambda_{\tens{Q}}$, an increase in $\lambda_{\tens{W}}$ leads to greater elongation of the cells in accordance with Eq.~\ref{eq:WErelation}. The role of the filament flow-aligning parameter $\lambda_{\tens{Q}}$ depends on the details of the active turbulent flow field which we now discuss in more detail.

\subsection*{Properties of the active turbulent flow field}
\begin{figure*}[ht] 
    \centering
    \includegraphics[width=\linewidth]{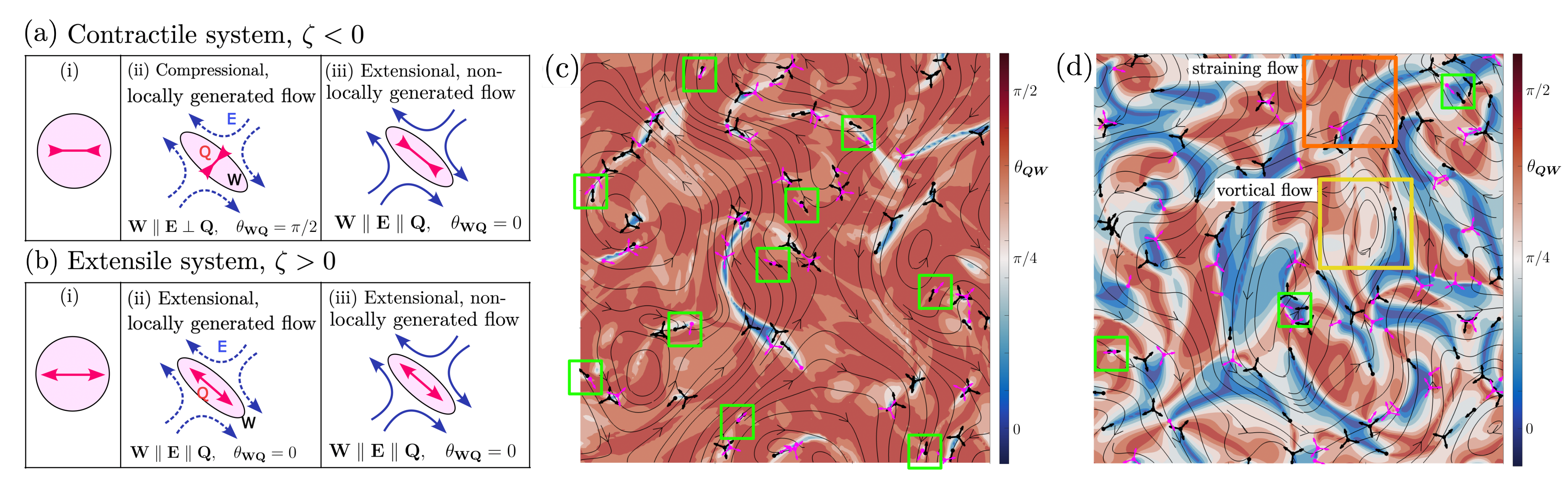}
  \caption{{\bf Elongated cells and active filaments orient in distinct ways in active turbulence.} (a) Contractile system: (i) isotropic cells (shaded circles) with contractile filaments (double-headed arrows) deformed by (ii) local compressional flows (dashed, blue arrows)) and (iii) non-local extensional flows (solid, blue arrows) of active turbulence. (b) Extensile system: Isotropic cells (shaded circles) with extensile filaments (double-headed arrows) deformed by (ii) local extensional flows (dashed, blue arrows) and (iii) non-local extensional flows (solid, blue arrows) of active turbulence. (c,d) Snapshots of the active turbulent domains in a contractile system ($\zeta < 0$) colored with the alignment angle $\theta_{\tens{W}\tens{Q}}$, the angle between the director of elongated cells and active filaments for (c) $\lambda_{\tens{Q}}=0$ and (d) $\lambda_{\tens{Q}}=0.2$. The continuous lines are velocity streamlines. The half integer defects in the nematic order of filaments and elongated cells are colored in black and magenta, respectively. Green outlines show examples of antiparallel configuration of $+1/2$ defects in shape and filaments which are close to each other. The orange and yellow boxes in (d) are, respectively, examples of straining and vortical flows.}\label{fig3fi}
\end{figure*}
Fig.~\ref{fig2fi}(a) plots the mean speed of the active turbulent flows for various values of filament flow-aligning parameter, comparing the contractile and the extensile cases. For $\lambda_{\tens{Q}}=0$, the velocities are similar in extensile and contractile systems, but they increase with $\lambda_{\tens{Q}}$ in an extensile system and decrease with $\lambda_{\tens{Q}}$ in a contractile system. To explain this we consider the different characteristics of extensile and contractile active flows.
Eq.~\ref{qdynamics} suggests that, to the leading order,
\begin{align}
 \partial_t \tens{Q} \sim \lambda_{\tens{Q}} \tens{E},
\label{eq:qapprox}
\end{align}
and, combining Eqs.~\ref{eq:EQrelation} and \ref{eq:qapprox}, we have  $\partial_t \tens{Q} \sim \frac{\zeta \lambda_{\tens{Q}}}{2\eta} \tens{Q}$. Thus, in an extensile system for which $\zeta > 0$, there is a bootstrapping effect, active flows enhance nematic order, leading to stronger flows emerging from  active hydrodynamic instabilities. On the other hand, contractile systems, for which $\zeta < 0$, oppose the build up of nematic order resulting in weaker active turbulent flows. While the nature of the active flow induced nematic ordering (or disruption) is dictated by whether the system is extensile or contractile (sign of $\zeta$), the strength of the effect is also influenced by the filament flow-aligning parameter $\lambda_{\tens{Q}}$. Thus, an increase in $\lambda_{\tens{Q}}$ increases the mean speed in extensile systems and decreases it in the contractile case.

Furthermore, the filament flow-aligning parameter affects the nature of the straining flows in active turbulence. An active turbulent velocity field consists of regions of extensional and compressional flows, which are distinguished locally based on the alignment of the strain rate tensor with respect to the orientation of the active filaments.  We define the flows as extensional when the extensional axis of the flow (the eigenvector of the strain rate tensor corresponding to the largest positive eigenvalue) is parallel to the axis of the active filaments i.e., $\tens{E} \parallel \tens{Q}$. Similarly, we define the flows as  compressional when the extensional axis of the flow is perpendicular to the axis of the filaments i.e., $\tens{E} \perp \tens{Q}$. Local extensional flows are characteristic of extensile filaments and local compressional flows are characteristic of contractile filaments.

Fig.~\ref{fig2fi}(b) and \ref{fig2fi}(c) plot $\psi(\theta_{\tens{E}\tens{Q}})$, the probability density function of the angle between the dominant eigenvector of the strain rate tensor and the director field of the active filaments in extensile and contractile systems respectively. When $\lambda_{\tens{Q}} = 0$, the distribution $\psi(\theta_{\tens{E}\tens{Q}})$ peaks at $ \theta_{\tens{E}\tens{Q}} = 0$ for extensile systems and at $ \theta_{\tens{E}\tens{Q}} = \pi/2$ for contractile systems, as expected. These distributions are nearly mirror images of
each other. However, the distributions and the symmetry in the distributions change with variations in filament flow-aligning parameter $\lambda_{\tens{Q}}$. With increase in $\lambda_{\tens{Q}}$, the distribution $\psi(\theta_{\tens{E}\tens{Q}})$ remains unimodal and the peak at $ \theta_{\tens{E}\tens{Q}} = 0$ becomes sharper for extensile filaments. For contractile filaments, increase in $\lambda_{\tens{Q}}$ weakens the peak at $ \theta_{\tens{E}\tens{Q}} = \pi/2$, the distribution becomes bimodal with the appearance of a new peak at $ \theta_{\tens{E}\tens{Q}} = 0$. To summarise, with increase in filament flow-aligning parameter, extensional flows get stronger in an extensile system while domains of extensional flows develop in an otherwise compressional contractile system. 

The existence of extensional flows in a contractile system disagrees with the approximate, local active-viscous stress balance anticipated from Eq.~\ref{eq:EQrelation}; instead they develop as a consequence of non-local hydrodynamics. A net straining flow is produced in a region due to viscous diffusion of momentum generated elsewhere in the field, a characteristic of Stokesian flows. In such straining flows, active filaments can rotate to align along the extensional axis producing regions of extensional flow, $\theta_{\tens{E}\tens{Q}} =0$. Since the mechanism, which again follows from Eq.~\ref{eq:qapprox}, is dependent on the filament flow-aligning parameter, the strength of the peak at $ \theta_{\tens{E}\tens{Q}} = 0$ increases with increase in $\lambda_{\tens{Q}}$.

On the other hand, non-locally generated flows do not give rise to such distinct and contrasting flow regions in extensile systems, since filaments already tend to orient along the extensional axis of local flows, and hence the peak at $ \theta_{\tens{E}\tens{Q}} = 0$ is only enhanced with increase in $\lambda_{\tens{Q}}$.

\subsection*{Alignment between active filaments and elongated cells}
We have confirmed in previous sections that the cells are predominantly deformed along the principal axis of the strain rate tensor, $\tens{W}\parallel\tens{E}$ and shown that, in contractile active turbulence, extensional flows may emerge causing the angle between $\tens{E}$ and the filament director $\tens{Q}$ to take a range of values between $0$ and $\pi/2$. We now discuss how this affects the angle between the orientation of the active filaments and the cell elongation axis, $\theta_{\tens{W}\tens{Q}}$.

Consider a system of isotropic cells (shaded circles) powered by co-occurring active filaments (two-headed arrows) shown schematically in column (i) in Fig.~\ref{fig3fi}(a). If the active filaments are contractile (shown by inward-pointing arrows), $\zeta < 0$ and active-viscous stress balance (Eq.~\ref{eq:EQrelation}) suggests that $\tens{E}\perp\tens{Q}$. That is, the active, straining fluid flows are compressional along the filament director axis, as expected for a contractile force dipole. But cells elongate along the extensional axis of the flow (in accordance with Eq.~\ref{eq:WErelation}) resulting in $\tens{W}\parallel\tens{E}\perp\tens{Q}$ (column (ii) in Fig.~\ref{fig3fi}(a)). 

However, we have argued that, as $\lambda_{\tens{Q}}$ increases, active turbulence can also produce effective extensional flows in contractile systems, i.e.~regions with $\tens{E}\parallel\tens{Q}$.  Isotropic cells deform along the extensional axis of the prevailing local flow as before, resulting in $\tens{W}\parallel\tens{E}\parallel\tens{Q}$ (column (iii) in Fig.~\ref{fig3fi}(a)).

 We highlight this behaviour in snapshots of contractile active turbulence in Figs.~\ref{fig3fi}(c) and~\ref{fig3fi}(d) where the domains are colored based on $\theta_{\tens{W}\tens{Q}}$, the angle between the director fields of the shape and of the active filaments. Regions with $\theta_{\tens{W}\tens{Q}} \approx \frac{\pi}{2}$ are colored red, and regions with  $\theta_{\tens{W}\tens{Q}} \approx 0$ are colored blue. 
 
 For $\lambda_{\tens{Q}}=0$ and contractile filaments Fig.~\ref{fig2fi}(c) indicates that $\tens{E}$ is typically perpendicular to $\tens{Q}$. Therefore case (ii) in Fig.~\ref{fig3fi}(a) holds, $\theta_{\tens{W}\tens{Q}} \approx \frac{\pi}{2}$ almost everywhere, and the domain in Fig.~\ref{fig3fi}(c) is almost entirely red. For $\lambda_{\tens{Q}}=0.2$, however Fig.~\ref{fig2fi}(c) indicates that a large fraction of filaments become elongated in the same direction as the extensional flows, $\tens{E}\parallel\tens{Q}$, corresponding to case (iii) in Fig.~\ref{fig3fi}(a). This is signalled by the blue patches in Fig.~\ref{fig3fi}(d), which we shall refer to as aligned patches.

The appearance of aligned patches is consistent with Monolayer Stress Microscopy measurements on conﬂuent MDCK layers \cite{nejad2024stress} which show that the area fraction of cells deformed in the direction of the active filaments (along the ﬁrst principal stress in the experiments) can be as large as 70\%. Regions that are strongly correlated with straining flows, $\theta_{\tens{W}\tens{Q}}\approx 0$ or $\theta_{\tens{W}\tens{Q}}\approx \pi/2$, tend to lie between vortices where extensional/compressional flows dominate vorticity (see Fig.~\ref{s2fig}). 

As expected, $\theta_{\tens{W}\tens{Q}}$ shows a qualitatively different behaviour with change in $\lambda_{\tens{Q}}$ in active turbulence driven by filaments with extensile activity, $\zeta > 0$. Here, the filaments are predominantly oriented along the extensional axis of the flow for all $\lambda_{\tens{Q}}$ as shown in Fig.~\ref{fig2fi}(b). Hence $\theta_{\tens{W}\tens{Q}} \approx 0$ almost everywhere.

\subsection*{Topological defects formed by elongated cells and active filaments}
\begin{figure}[t]
\includegraphics[width=\columnwidth]{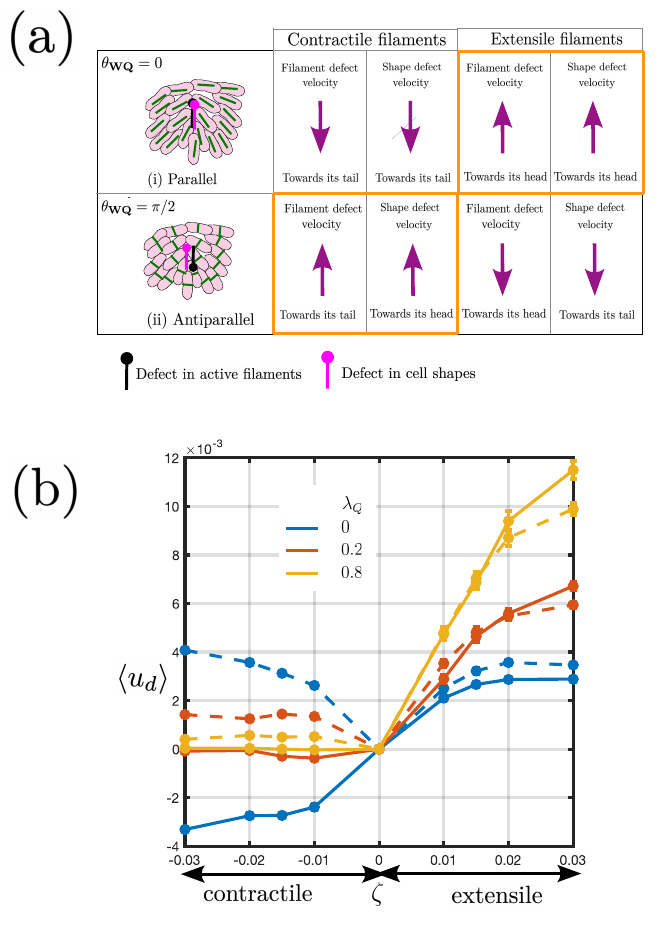}
 \caption{{\bf Topological defects in the filament and shape fields}: (a) Schematic illustrating the parallel (top) and antiparallel (bottom) configurations of co-localised $+\frac{1}{2}$ topological defects formed in regions where the cells and filaments are aligned ($\theta_{\tens{W}\tens{Q}}\approx 0$) or misaligned ($\theta_{\tens{W}\tens{Q}}\approx \frac{\pi}{2}$). Elongated cells are colored in pink, active filaments in green. The orientation of the shape and filament defects are indicated in magenta and black colors, respectively. Columns 2 and 3 indicate the velocity direction of the defects in contractile and extensile systems, respectively. Filament defects move towards their tails (heads) in contractile (extensile) systems, shape defects move with the filament defects. Boxes with orange outlines highlight scenarios predominantly observed in extensile and contractile systems. (b) Average flow velocity  $\langle u_d \rangle$ measured at the center of $+\frac{1}{2}$ shape and filament defects in extensile and contractile systems. $\langle u_d \rangle$ is defined as $+$ve towards the head of the defect. The dashed (solid) lines correspond to defects in cell shape (filaments).}
\label{fig4fi}
\end{figure}
Motile, long-lived $\pm\frac{1}{2}$ defects, which contribute to generating active flows, are characteristic of 2D active turbulence \cite{doostmohammadi2018active, giomi2014defect,serra2023defect}. However, in the study and analysis of defects in cellular layers \cite{saw2017topological, zhang2022topological, ardavseva2022topological, kawaguchi2017topological} no distinction has yet been made between defects in the active filament and cell shape director fields. Our model allows us to make this distinction, and we find that defects are created in both the filament and the shape fields. 

Figs.~\ref{fig3fi}(c) and~\ref{fig3fi}(d)  illustrate the position and orientation of filament and shape defects in a contractile system for $\lambda_{\tens{Q}}=0$ and $\lambda_{\tens{Q}}=0.2$ respectively. The defects tend to lie on the boundaries between aligned and misaligned patches, as observed in the measurements on epithelial monolayers \cite{nejad2024stress}: this is natural as defects are locations with large changes in the orientation field. 
The figures show that the positions of the defects in shape and filaments are often correlated. However, even if the filament and shape defects are formed at the same location, they can take different relative orientations; the limiting cases for $+\frac{1}{2}$ defects are illustrated in Fig.~\ref{fig4fi}(a). If the elongated cells and active filaments are aligned ($\theta_{\tens{W}\tens{Q}}\approx 0$), the $+\frac{1}{2}$ filament and shape defects point in the same direction.  However, if the elongated cells and active filaments are orthogonal ($\theta_{\tens{W}\tens{Q}}\approx \pi/2$) the respective $+\frac{1}{2}$ defects take an antiparallel configuration, pointing in opposite directions. In our simulations, we more often find co-localised filament and shape defects in a parallel configuration in extensile systems and in an antiparallel configuration in contractile systems. 

The asymmetry in the structure of a $+\frac{1}{2}$ topological defect formed by active filaments results in self-propulsion of the defect.  The propulsion velocity is proportional to the activity of the filaments and is inversely proportional to the viscosity of the medium \cite{ronning2022flow}. In a contractile system the filament defects move towards their tails as is well 
known from single order parameter nematohydrodynamic theories \cite{doostmohammadi2018active}. 
On the other hand, $+\frac{1}{2}$ topological defects formed by the elongated cells do not self-propel, instead they are advected by the active turbulent flow field, often moving along with the associated filament $+\frac{1}{2}$ defect. Thus, as a consequence of the relative orientation of the filament and shape defects, in an aligned region the shape defects will also move towards their tails, whereas in a region with $\theta_{\tens{W}\tens{Q}}\approx \pi/2$ they will move towards their heads. This is illustrated in Fig.~\ref{fig4fi}(a) and provides an explanation for the heretofore puzzling observation that topological defects in confluent cell layers are observed to move either towards their heads or towards their tails even though the cell filaments exert contractile forces.

The similarities and disparities arising in the dynamics of shape and filament defects can be illustrated from the simulation data by measuring the average fluid velocity at the center of $+\frac{1}{2}$ defects in an active turbulent flow field. This is plotted in Fig.~\ref{fig4fi}(b). The velocity is defined positive in the direction of orientation of the $+\frac{1}{2}$ defect, from tail to head. The magnitude of measured velocity increases with increasing activity for both shape and filament defects irrespective of whether the filaments are extensile or contractile. Moreover, as expected, the velocities at the center of shape and filament defects are comparable for extensile systems, and they increase with increasing filament flow-aligning parameter.

For contractile systems, since the change in filament flow-aligning parameter can change the distribution of $\theta_{\tens{E}\tens{Q}}$   (Fig.~\ref{fig2fi}(c)), the difference in the apparent velocity measured at the center of filament and shape defects is also dependent on $\lambda_{\tens{Q}}$. For $\lambda_{\tens{Q}}=0$, $\theta_{\tens{E}\tens{Q}}$ peaks at $\approx \frac{\pi}{2}$. The shape and filament defects almost always move together in an antiparallel configuration, causing their velocities to be comparable in magnitude but oriented in opposite directions. This correlation reduces with an increase in the value of $\lambda_{\tens{Q}}$ because of the decrease in the strength of active turbulent flow (Fig. \ref{fig2fi}(a)) and because the shape and filament defects become less correlated in position and orientation (Fig.~\ref{fig2fi}(c)).

\subsection*{Summary}
Using two tensor order parameters to differentiate the role of the force-generating active filament machinery in cells from the cells' shapes themselves, we construct a continuum theory to describe the dynamics of cellular layers and tissues. The theory shows that straining active flows can elongate isotropic cells, which form nematic domains on the size of the active length scale. We discuss the properties of the active turbulent flow field emphasising the relevance of a rheological quantity, the flow-aligning parameter of the filaments. As this increases, flow alignment of active filaments becomes more prevalent, and regions where the filament director field is parallel and regions where it is perpendicular to the straining flow can co-exist in contractile active nematics. This agrees with the experimental observations of patches of alignment and misalignment between cell elongation and the direction of active stress in epithelial MDCK layers.

We distinguish topological defects generated in the filament director field, and in the cell elongation director field. In the former the defects move towards their tails in a contractile system as is well known from standard active nematic theories. Shape defects tend to be associated with filament defects, and move with them but point in a parallel (antiparallel) direction in aligned (misaligned) regions.  Hence our approach explains why the shape defects in contractile cell monolayers are observed to move either towards their tails or their heads.

Our theory highlights the importance of distinguishing the roles of active filaments and cell shape, and reconciles the idea of contractile cells exhibiting extensile behaviour, thus advancing the continuum modelling of cellular collectives. It also poses questions and suggests directions for new experiments on cell monolayers. We have referred to stress fibers and cytoskeletal force generating machinery as active filaments, but details of how these generate dipolar forces is unclear. In particular, it would be interesting to simultaneously measure filament and cell shape orientations to test the ideas presented here. Filament dynamics is difficult to follow, but may be accessible by perturbing monolayers by, e.g., stretching. Moreover, measuring the velocity distribution of cells and of defects in the shape field, in experiments under varying conditions with different cell-lines, will help to identify the physical regions of parameter space in the continuum model, and to identify any additional relevant physics.

\subsection*{Supporting Information Appendix (SI)}
\setcounter{secnumdepth}{2}
\renewcommand\thesubsection{S\thesection\arabic{subsection}}
\setcounter{figure}{0}
\renewcommand{\thefigure}{S\arabic{figure}}

\subsection{Temporal autocorrelation of the shape tensor}
\label{section:autocorr}
\begin{figure}[b]
\includegraphics[width=0.7\columnwidth]{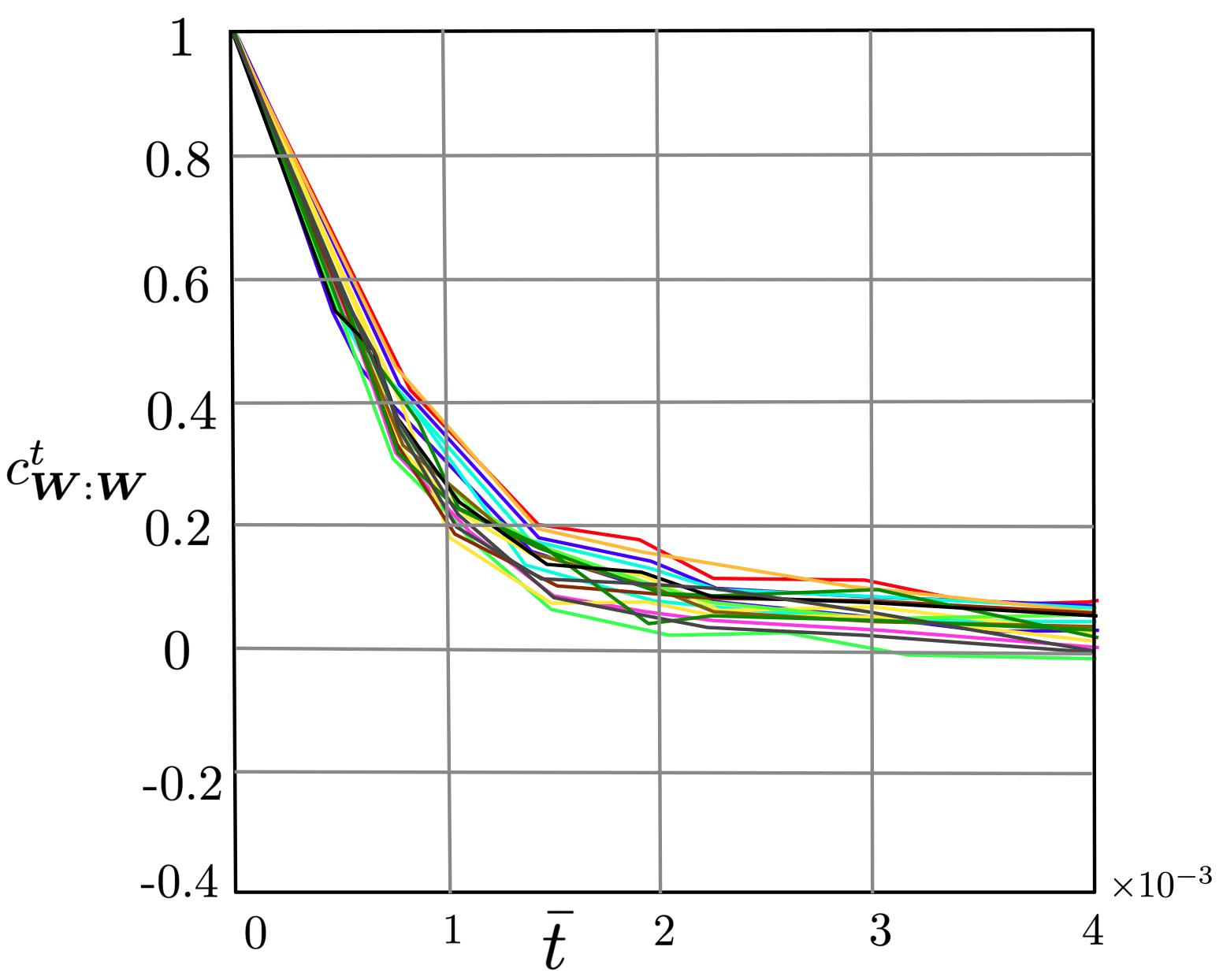}
    \caption{Autocorrelation of the shape tensor $c_{\tens{W}:\tens{W}}^t$ as a function of rescaled time $\bar{t}= t \zeta/\eta$ for the same parameters as in Fig.~\ref{fig1fi}(c). }
         \label{fig:timecorr}
\end{figure}

\begin{figure}[t]
\includegraphics[width=0.7\columnwidth]{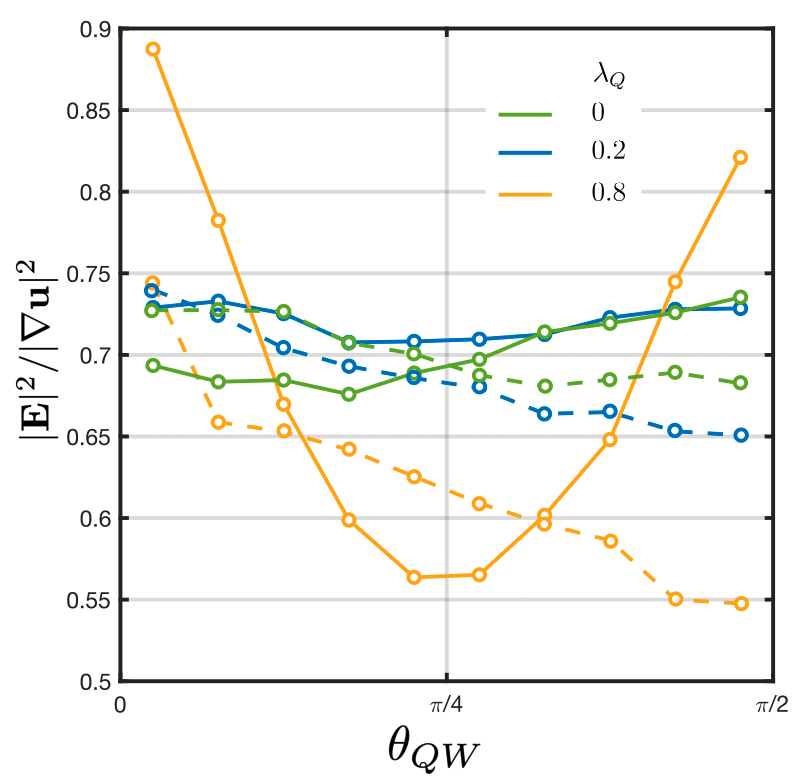}
    \caption{The ratio of the straining flow to the total velocity gradient $|\tens{E}|^2/|\boldsymbol{\nabla} \vec{u}|^2$ as a function of misalignment between filaments and cell shape $\theta_{\tens{Q}\tens{W}}$. The dashed (solid) lines show systems with extensile (contractile) filaments. The parameter values are the same as in Fig.~\ref{fig2fi}(b) and (c).} 
         \label{s2fig}
\end{figure}

The time scale of active flows, $\tau_a \sim \eta/\zeta$ can be obtained from the balance between the active stress and the viscous stress (Eq.~\ref{eq:EQrelation} in the manuscript). Since the flows result in the deformation of cells, we calculated the autocorrelation of the shape tensor to determine the persistence time of the deformed cells. The autocorrelation function is calculated as $c_{\tens{W}:\tens{W}}^t = \langle \tens{W}(r,t_0+t) : \tens{W}(r,t_0)\rangle_{r,t_0}/\langle \tens{W}(r,t_0) : \tens{W}(r,t_0)\rangle_{r,t_0}$, where the average $\langle \rangle_{r,t_0}$ is defined over space $r$ and time $t_0$,  and plotted in Fig.~\ref{fig:timecorr} for the same range of parameters as in Fig.~\ref{fig1fi}(c) in the manuscript. As evident from the figure, the decorrelation time of the shape tensor is approximately correlated with the active time scale, $\tau_a$.
\subsection{Stronger extensional flows in aligned and misaligned regions}
\label{section:ebygradu}
In order to show the effect of straining flows in determining the regions of alignment and misalignment between the active filaments and elongated cells we plot the ratio of the straining flow to the total velocity gradient $|\tens{E}|^2/|\boldsymbol{\nabla} \vec{u}|^2$ as a function of misalignment angle between filaments and cell shape $\theta_{\tens{Q}\tens{W}}$. The dashed (solid) line shows a system with extensile (contractile) filaments. It may be seen that the distribution is monotonic in extensile systems irrespective of the value of $\lambda_{\tens{Q}}$. However, for large $\lambda_Q$ in contractile systems, the regions of alignment ($\theta_{\tens{Q}\tens{W}} \approx 0$) and misalignment ($\theta_{\tens{Q}\tens{W}} \approx \frac{\pi}{2}$) correspond to regions with relatively larger contribution of rate of strain tensor to the velocity gradient.

\subsubsection*{Movies}
\textbf{Movie 1}: Dynamics of a system with contractile active filaments, with filament flow aligning parameter $\lambda_Q=0$. Left panel: cell shape orientation and defects in the cell shape field. The background color shows the magnitude of the cell elongation $|\tens{W}|$. Right panel: active flows are shown as streamlines, defects in cell shape (filaments) are shown in magenta (black). The background colour shows the misalignment angle $\theta_{\tens{Q}\tens{W}}$. For a zero filament shear alignment, the cell shape is almost everywhere perpendicular to the the filament orientation.\\

\textbf{Movie 2}: Dynamics of a system with contractile active filaments, with filament flow aligning parameter $\lambda_Q=0.8$. Left panel: cell shape orientation and defects in the cell shape field. The background color shows the magnitude of the cell elongation $|\tens{W}|$. Right panel: active flows are shown as streamlines, defects in cell shape (filaments) are shown in magenta (black). The background color shows the misalignment angle $\theta_{\tens{Q}\tens{W}}$. For a non-zero filament shear alignment, the monolayer forms regions where cell shape is aligned with the filament orientation.\\

\subsubsection*{Acknowledgment}
We thank Jacob Notbohm for helpful discussions. SPT would like to thank the Royal Society and the Wolfson Foundation for a Royal Society Wolfson Fellowship
award and acknowledges the support by the Department of Science and Technology, India, via the research grant CRG/2023/000169.
JMY acknowledges support from the UK EPSRC (Award EP/W023849/1) and ERC Advanced Grant ActBio (funded as UKRI Frontier Research Grant
EP/Y033981/1). This research was supported in part by grant no. NSF PHY-2309135 to the Kavli Institute for Theoretical Physics (KITP).

\bibliography{references}
\end{document}